\documentclass[twocolumn,prl,superscriptaddress]{revtex4}

\usepackage{graphicx}
\usepackage{amsmath}
\usepackage{color}
\usepackage{units}
\usepackage[latin1]{inputenc}

\begin{document}

\title{Observation of one-way Einstein-Podolsky-Rosen steering}

\author{Vitus H\"{a}ndchen}
\author{Tobias Eberle}
\author{Sebastian Steinlechner}
\author{Aiko Samblowski}
\affiliation{Centre for Quantum Engineering and
Space-Time Research - QUEST, Leibniz Universit\"at Hannover,
Welfengarten 1, 30167 Hannover, Germany}
\affiliation{Max-Planck-Institut f\"ur Gravitationsphysik
(Albert-Einstein-Institut) and\\ Institut f\"ur Gravitationsphysik
der Leibniz Universit\"at Hannover, Callinstra{\ss}e 38, 30167 Hannover,
Germany}
\author{Torsten Franz}
\author{Reinhard F.\,Werner}
\affiliation{Centre for Quantum Engineering and
Space-Time Research - QUEST, Leibniz Universit\"at Hannover,
Welfengarten 1, 30167 Hannover, Germany}\affiliation{Institut f\"ur Theoretische Physik der Leibniz Universit\"at Hannover, Appelstra{\ss}e 2, 30167 Hannover, Germany}
\author{Roman Schnabel}
\email[corresponding author: ]{roman.schnabel@aei.mpg.de}
\affiliation{Centre for Quantum Engineering and
Space-Time Research - QUEST, Leibniz Universit\"at Hannover,
Welfengarten 1, 30167 Hannover, Germany}\affiliation{Max-Planck-Institut f\"ur Gravitationsphysik
(Albert-Einstein-Institut) and\\ Institut f\"ur Gravitationsphysik
der Leibniz Universit\"at Hannover, Callinstra{\ss}e 38, 30167 Hannover,
Germany}

\maketitle

\textbf{The distinctive non-classical features of quantum physics were first discussed in the seminal paper \cite{EPR1935} by A.\,Einstein, B.\,Podolsky and N.\,Rosen (EPR) in 1935. In his immediate response~\cite{Schroedinger1935} E.\,Schr\"odinger introduced the notion of \emph{entanglement}, now seen as the essential resource in quantum information \cite{Ekert91,Zeilinger00,Braunstein2005} as well as in quantum metrology \cite{AAS10, SMML10, SqzNatPhys11}. Furthermore he showed that at the core of the EPR argument is a phenomenon which he called \emph{steering}. In contrast to entanglement and violations of Bell's inequalities, steering implies a direction between the parties involved. Recent theoretical works have precisely defined this property  %and the question arose wether there are bipartite states showing steering only in one direction 
\cite{Wiseman2007, Wiseman2009}. Here we present an experimental realization of two entangled Gaussian modes of light by which in fact one party can steer the other but not conversely. The generated one-way steering gives a new insight into quantum physics and may open a new field of applications in quantum information.}

Steering can be described considering two remote experimenters, Alice and Bob, who share a bipartite quantum state. Their local systems are in a mixed state and should, therefore, permit a decomposition into pure states. Schr\"odinger found that within quantum mechanics certain states do not allow such decomposition locally. Depending on the observable Alice chooses to measure, Bob's local state is decomposed into incompatible mixtures of conditional states. So if pure states were a local complete description of Bob's system, this would require some interaction from Alice to Bob. This is what Schr\"odinger named steering and Einstein later called the ``spooky action at a distance''. The first experimental demonstration of this effect was achieved by Ou \emph{et al.} \cite{Ou1992}, followed by a greater number of experiments as reviewed in \cite{Reid2009}.

\begin{figure}[htb]
  \center
  \includegraphics[width=6cm]{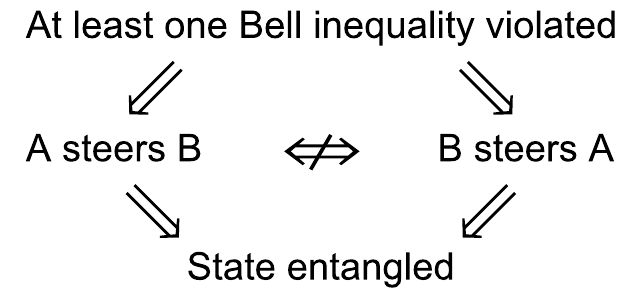}
  \caption{
\textbf{Implications of inseparability criteria.}\,A violation of at least one Bell inequality implies steering in both directions. If steering is only present in one direction, no Bell inequality can be violated. But any certification of steering implies that the state is entangled. The converse implications are not true: Entangled states are not necessarily steering states and steering does not imply the violation of a Bell inequality.
}
  \label{fig: implications}
\end{figure}

Steering is strictly stronger than entanglement, but does not entail a violation of a Bell inequality \cite{Werner1989}. In contrast to the latter two, Alice and Bob have certain roles in the steering scenario which are not interchangeable. This intrinsic asymmetry leads to implications of inseparability criteria as shown in Fig.~\ref{fig: implications}. The question arises \cite{Wiseman2007} whether there are physical states certifying steering only in one direction for arbitrary observables. This \textit{one-way} steering would lead to the peculiar situation that two experimenters measuring the same observables on their subsystems would describe the same shared state in qualitatively different ways. Where in general this question can not be answered yet, in the regime of Gaussian states and homodyne measurements the answer is yes. 
In a pioneering paper by H.-A.\,Bachor and co-workers, two-way steering with an asymmetry in the steering strengths was 
observed~\cite{Wagner2008}. Their theoretical analysis proposes a possible extension of their setup towards observing one-way steering. In a more recent theoretical work, an intra-cavity nonlinear coupler was proposed to generate Gaussian one-way steering \cite{Midgley2010}.

Here we propose and experimentally certify the realization of Gaussian one-way steering with two-mode squeezed states. Our states were generated by first superimposing a squeezed mode with a vacuum mode at a balanced beam splitter. By introducing additional  amounts of vacuum to Bob's mode the overall state's asymmetry was stepwise driven through the one-way regime, finally loosing all steering properties. The most significant one-way states were qualified by the Reid criterion giving $0.908\pm0.003$ for the direction from Alice to Bob and $1.206\pm0.004$ for the reverse direction, where the normalization was chosen such that below $1$ steering is certified.

\begin{figure*}[htb]
\centering
\includegraphics[width=17cm]{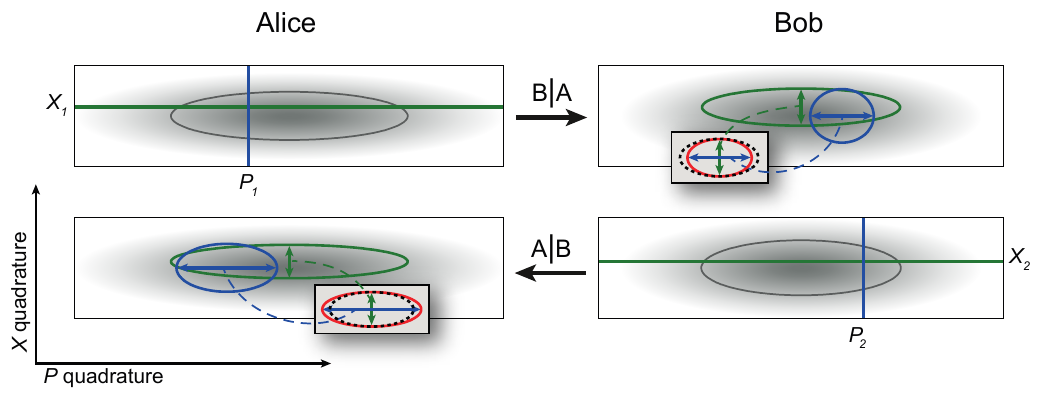}
  \caption{\textbf{Gaussian one-way EPR-steering visualized in phase space.}\,The local Wigner functions of a bipartite quantum state, as presented in this work, are represented by the grey ellipses as well as by the background clouds. The axes are not to scale. In the upper panels steering from Alice to Bob is shown. Two exemplary measurement results $X_1$ and $P_1$ are depicted by the green and blue line. The appertaining conditional states at Bob's system are depicted by the accordingly colored ellipses, with their uncertainties in the respective quadratures given by the arrows. Any hypothetical common refinement, depicted by the red ellipse in the inlay, may not exceed these uncertainties. Hence, it would violate the Heisenberg Uncertainty Relation, shown in black. In the lower panels a non-steering situation from Bob to Alice is shown. In this case a common refinement is possible, i.e.\ the uncertainty relation is not violated.}
  \label{fig: epr-steering}
\end{figure*}

To analyze the steering scenario, we start with the bipartite situation in which Alice sends quantum states to Bob. If Bob locally observes a mixed state, this can be decomposed into convex combinations of purer states. These decompositions can be seen as more precise descriptions of his system. Indeed, any information that Alice has about the state will give a decomposition into conditional states. This can be seen in the upper panels of Fig.~\ref{fig: epr-steering} for the case of a Gaussian system and quadrature measurements. Two exemplary measurement results $X_1$ and $P_1$ which Alice obtains on her system are depicted by the green and blue line. The related conditional states on Bob's side are shown by the accordingly colored ellipses. For all measurement results Alice can obtain, these ellipses will have the same shape and just their position in phase space will be different. So Alice's $X$- and $P$-results give two different decompositions of Bob's system.

The argument by EPR and Schr\"odinger is now, that measurements on Alice's side should not influence Bob's system. So the decomposition of Bob's state should be independent of Alice's choice of observable. This implies, that the conditional decompositions, which depend on Alice's choice, should have a common refinement, which does not depend on Alice's choice. This refinement should show an $X$-uncertainty that is at most as large as the one of Bob's $X$-conditional state (green arrow). At the same time, it should show a $P$-uncertainty that is at most as large as the one of Bob's $P$-conditional state (blue arrow). We have depicted this hypothetical state in the inlay as red ellipse. But this state would clearly violate the Heisenberg uncertainty relation, depicted by the black dotted ellipse.

The absence of a common refinement leads to the conclusion that Alice's choice of observable somehow changes the states of Bob's system, which Schr\"odinger called steering. More formally, we define a bipartite state to be steerable with respect to Alice's observables, if the resulting conditional state decompositions of Bob's state do not allow a common refinement. We say that the state is steerable from Alice to Bob, if there are some observables for which it is steerable. This description of steering is close to Schr\"odingers original presentation. It is equivalent to a modern definition based on the existence of certain classical models as given in the seminal paper \cite{Wiseman2007}.

Since we consider only Gaussian states and homodyne measurements, our description of steering is equivalent to a definition by M.\,Reid~\cite{Reid1989,Wiseman2009}. Her definition is based on Heisenberg Uncertainty Relations for conditional measurements of the amplitude and phase quadrature $X$ and $P$ of light fields. A state is steerable from Alice to Bob, if the following conditional Heisenberg Uncertainty Relation is violated:
\begin{equation}
\text{V}_{B|A}(X_B) \cdot \text{V}_{B|A}(P_B)\geq 1\ .
\label{eq: EPR-Reid1}
\end{equation}
Here, $\text{V}_{B|A}(X_B)$ denotes the conditional variance of $X_B$, i.e.\ the variance of Bob's measurements conditioned on Alice's results. We have chosen the units such that the right hand side is $1$. A violation of this inequality is exactly what is shown in the upper inlay of Fig.~\ref{fig: epr-steering} where the red ellipse is smaller than the black.

Conversely, steering from Bob to Alice is certified if the following inequality is violated:
\begin{equation}
\text{V}_{A|B}(X_A) \cdot \text{V}_{A|B}(P_A)\geq 1\ .
\label{eq: EPR-Reid2}
\end{equation}
This converse scenario is shown in the lower panels of Fig.~\ref{fig: epr-steering} for the same quantum state as in the upper panels. Both measurement results which Bob obtains give related conditional states on Alice's side and permit two different decompositions. But this time these conditional decompositions do have a common refinement that does not violate the uncertainty relation. So, in terms of Schr\"odinger, Bob's measurements do not steer Alice's system, as an underlying description with pure states is possible. Therefore, the state analyzed in Fig.~\ref{fig: epr-steering} shows one-way steering in the Gaussian regime.

\begin{figure}[htb]
  \center
  \includegraphics[width=8cm]{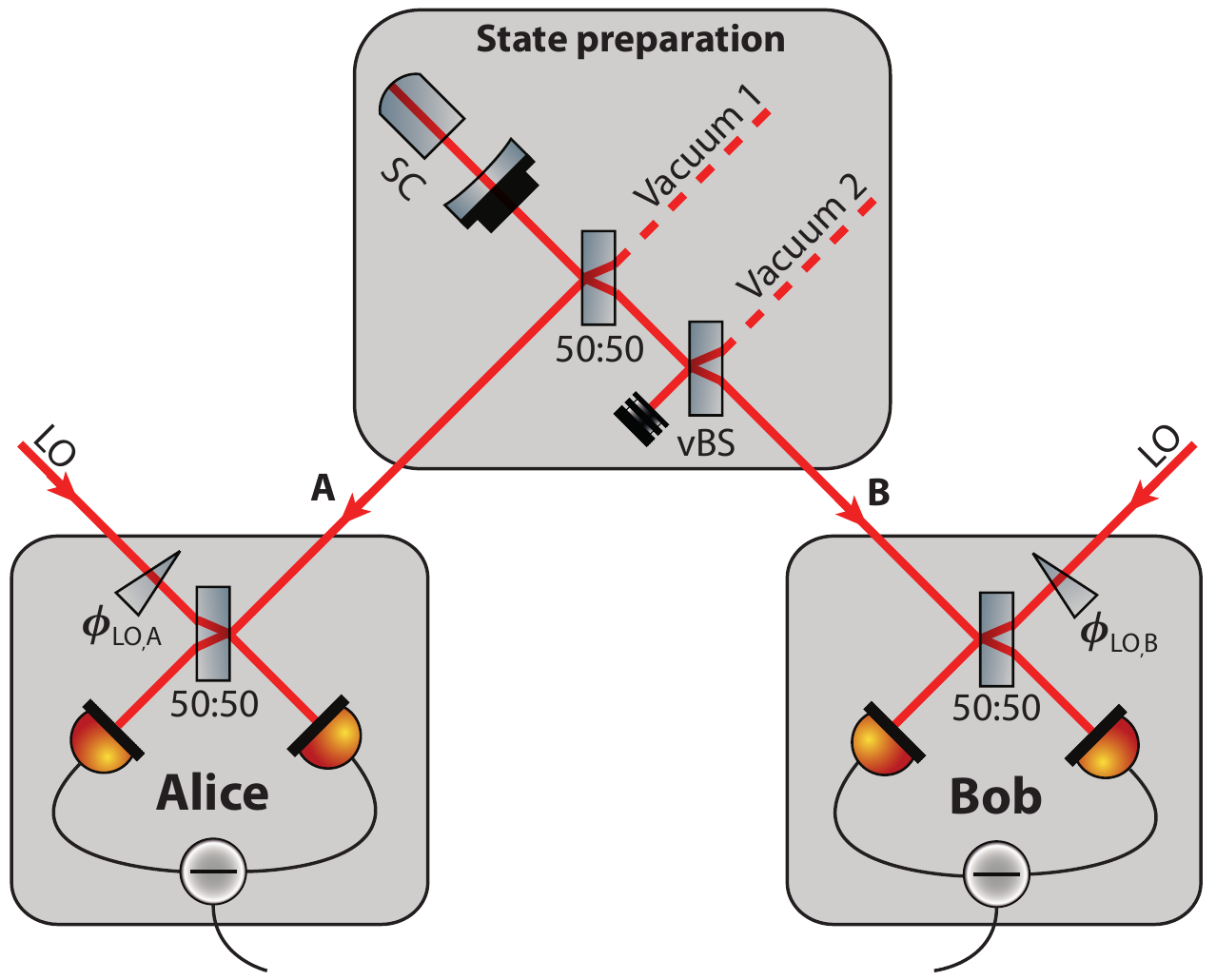}
  \caption{
 {\bf Schematic of the experimental setup.} \,A squeezed-light field at $1550$\,nm, produced by a squeezing cavity (SC), is superimposed at a balanced beam splitter with a vacuum mode. A variable beam splitter (vBS) is implemented in one output mode to change the contribution of a second vacuum mode. Measurements are performed by balanced homodyne detection where the measured quadrature is chosen by the phase $\phi_{LO}$ of the local oscillator (LO).
}
  \label{fig: ExperimentalSetup}
\end{figure}

The experimental setup we used to generate these one-way steering states is schematically shown in Figure \ref{fig: ExperimentalSetup}. The continuous-wave $10.2$\,dB squeezed state at $1550$\,nm was generated by type I parametric down conversion in a half-monolithic cavity. After the superposition with vacuum on a first balanced beam splitter, output mode B was sent through a half-waveplate and a polarizing beam splitter. This setup allowed the preparation of mode B with an adjustable contribution of a second vacuum mode. The measurements at A and B were performed by balanced homodyne detection. Both detectors could independently choose the measured quadrature by adjusting the phase of their local oscillators. The signals of the homodyne detectors were simultaneously recorded with a data acquisition system. A more detailed description of the squeezed-light source, the homodyne measurement setup and the locking scheme is given in \cite{Eberle2011b}.

\begin{figure}[htb]
  \center
  \includegraphics[width=8.4cm]{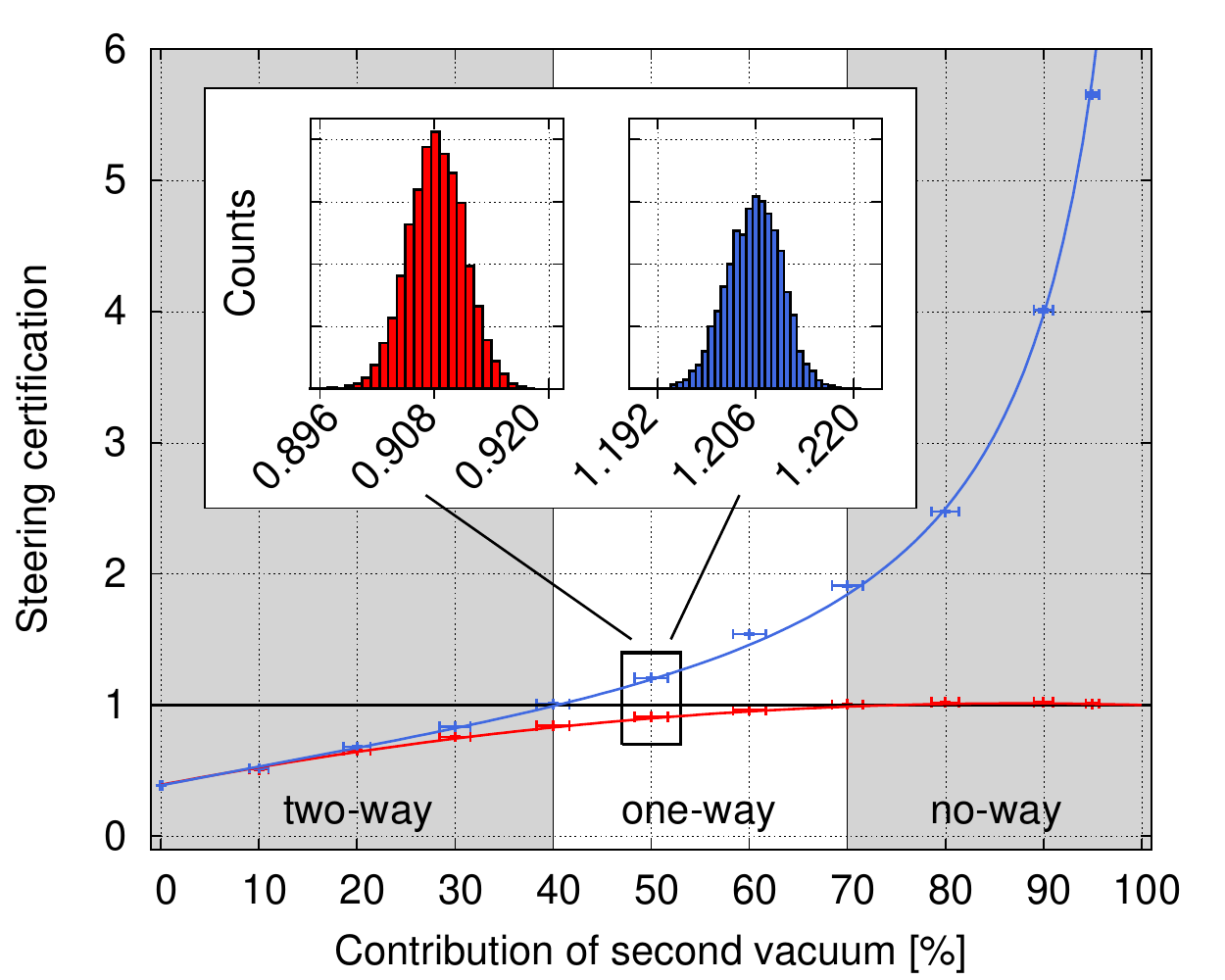}
  \caption{
{\bf Certification of one-way steering.} \,Shown are measurement results of the conditional variance products according to the criteria \,(\ref{eq: EPR-Reid1}) and (\ref{eq: EPR-Reid2}) versus an increasing contribution of the second vacuum mode in mode B. One-way steering is observed if the value for one steering direction is below unity, whereas the value for the other steering direction is above the unity benchmark. This is fullfilled in the white region and most significantly at a vacuum contribution of $50\%$, as shown by the measurement histograms.
}
  \label{fig: epr_vs_loss}
\end{figure}

Figure \ref{fig: epr_vs_loss} shows the main result of this work. The conditional variance products from inequalities\,(\ref{eq: EPR-Reid1}) and (\ref{eq: EPR-Reid2}) for Alice's ability to steer Bob (lower line, red) and Bob's ability to steer Alice (upper line, blue) are plotted against the contribution of the second vacuum mode. For values between $0\%$ and $95\%$ we performed a partial tomographic measurement. In order to determine the means and standard deviations of the conditional variance products a bootstrapping method was used. For $10^4$ times we randomly chose $10^6$ data points out of the total $5 \cdot 10^6$. From these we calculated the two conditional variance products for each data set. A histogram of these values for $50\%$ vacuum contribution is shown in the small boxes in Fig.\,\ref{fig: epr_vs_loss}. This is the setting where the observed one-way steering effect becomes most obvious. For Alice (left box) the mean of $0.908$ is $31$ standard deviations below $1$ whereas for Bob (right box) the mean of $1.206$ is $53$ standard deviations above $1$.

The two solid lines in Fig.\,\ref{fig: epr_vs_loss} are theory curves taking into account the optical detection efficiencies and the parameters of the squeezed-light source. For a vacuum contribution smaller than $39\%$, both Alice and Bob can steer the respective remote subsystem, whereas for a contribution larger than $70\%$ neither of them can. One-way steering is observed precisely between these two values in the white region in Fig.~\ref{fig: epr_vs_loss}.

While for our present experiment one of the output modes of the variable beam splitter was dumped, a tripartite situation arises when instead a third party, Charlie, receives this mode. For symmetry reasons Alice would then also be able to steer Charlie, in fact, simultaneously to Bob. We can further say, that neither Bob can steer Charlie nor conversely since the input of the second beam splitter already has a vacuum mode contribution of $50\%$ due to the first beam splitter. Steering in the presence of just one squeezed mode is only possible for vacuum contributions less than $33\%$~\cite{Eberle2011a}.

In conclusion, our experimental scheme provided the generation of Gaussian one-way steering with high significance. The criterion (\ref{eq: EPR-Reid1}) for steering from Alice to Bob was violated by more than 30 standard deviations whereas criterion (\ref{eq: EPR-Reid2}) for steering from Bob to Alice was \emph{not} violated with a significance of more than 50 standard deviations. Hence, depending on whether Alice tries to steer Bob's system, or Bob tries to steer Alice's system, our prepared state provided two opposing ans\-wers. This one-way property of EPR steering gives a new insight into the counterintuitive nature of quantum physics. It may have applications in bipartite and multipartite quantum key distribution and in information science in general. Its full nature, however, is up to now basically unexplored.

\begin{acknowledgments}
We thank J.\,Duhme for helpful discussions. This research was supported by the EU FP\,7 project Q-ESSENCE (Grant agreement number 248095). VH, TE, SS and AS thank the IMPRS on Gravitational Wave Astronomy for support. TF and RFW acknowledge support by the EU FP\,7 project COQUIT (Grant agreement number 233747) and BMBF project QuoRep.
\end{acknowledgments}

\end{document}